\documentclass[12pt,aps,prd,showpacs,amsmath,amssymb]{revtex4}
\input epsf
\begin{document}
\title{Exclusive semileptonic $B_s$ decays to excited $D_s$ mesons:\\Search of
$D_{sJ}(2317)$ and $D_{sJ}(2460)$}
\author{Ming-Qiu Huang}
\affiliation{Department of Applied Physics, Nat'l University of Defense
Technology, Hunan 410073, China}
\date{\today}
\begin{abstract}
We study the exclusive semileptonic decays $B_s\to D_{s0}^*\ell\bar\nu$ and
$B_s\to D_{s1}^*\ell\bar\nu$, where $p$-wave excited $D_{s0}^*$ and $D_{s1}^*$
states are identified with the newly observed $D_{sJ}(2317)$ and $D_{sJ}(2460)$
states. Within the framework of HQET the Isgur-Wise functions up to the
subleading order of the heavy quark expansion are calculated by QCD sum rules.
The decay rates and branching ratios are computed with the inclusion of the order
of $1/m_Q$ corrections. We point out that the investigation of the $B_s$
semileptonic decays to excited $D_s$ mesons may provide some information about
the nature of the new $D_{sJ}^*$ mesons.
\end{abstract}
 \maketitle

\section{Introduction}
\label{sec1}

Recently BaBar collaboration reported a narrow state with $J^P=0^+$ at a rather
low mass $2317$ MeV decaying to $D_{s} \pi^{0}$ \cite{Babar}. This observation
was quickly confirmed by CLEO, and another narrow state with $J^P=1^+$ at $2457$
MeV was found in the $D^{*}_{s} \pi^0$ channel \cite{Cleo}. Both states have been
confirmed by Belle \cite{Belle}. Since these states lie below $D K$ or $D^\ast K$
threshold, they decay instead via isospin-violating transitions.

These newly observed states have attracted much attention because their measured
masses and widths do not match the predictions from potential-based quark models
\cite{GI85}. To resolve the discrepancy, many theoretical speculations have
appeared in the literature.  Bardeen $et$ $al.$ interpreted them as
positive-parity $\bar c s$ $(0^+, 1^+)$ spin doublet of the $D_s$ and $D_s^*$
negative-parity $c\bar s$ $(0^-, 1^-)$ ground states in the framework of chiral
symmetry \cite{bardeen}. Based on the quark-antiquark picture, various
theoretical models are modified to accommodate the low masses and the narrow
widths for the new states \cite{colangelo,cahn,godfrey,datta,hnli}. QCD sum rule
analysis in \cite{DHLZ} supports the quark-antiquark postulation, both new
states, $D_{sJ}(2317)$ and $D_{sJ}(2460)$, are identified to be the $\bar c s$
excited $0^+$ and $1^+$ states in the $j_l={1\over 2}^+$ doublet. Apart from the
quark-antiquark interpretation, the $D^{*}_{sJ}(2317)$ meson has been interpreted
as a $DK$ molecule \cite{BCL}, a $D_s\pi$ molecule \cite{Szcz}, a four-quark
state \cite{CHH}, and a mixing of the conventional state and the four-quark state
\cite{BPP}.

Motivated by the interpretation that the two new states are the excited $p$-wave
$D_s$ states belonging to the $(0^+, 1^+)$ doublet with $j_l={1\over 2}^+$ in
\cite{DHLZ}, it is worthwhile to investigate the $B_s$ exclusive semileptonic
decays to this doublet, ie. $B_s\to D_{s0}^*\ell\bar\nu$ and $B_s\to
D_{s1}^*\ell\bar\nu$. The search of these decay modes may provide some
information to understand the nature of these new states and to clarify the
controversy. Although current $B$ factories do not produce $B_s$ meson, these
decays can be studied at future hadron $B$ factories.

In this paper we shall use QCD sum rules \cite{svzsum,hqetsum} in the framework
of the heavy quark effective theory (HQET) \cite{HQET,neubert1} to study
exclusive semileptonic $B_s$ decays to excited spin symmetry doublet
$(D_{s0}^*,D_{s1}^*)$ mesons. HQET is a useful tool to describe the spectroscopy
and weak decays of hadrons containing a single heavy quark, it provides a
systematic method to compute the properties of heavy hadrons via the $1/m_Q$
expansion, where $m_Q$ is the heavy quark mass. The study for the non-strange $B$
semileptonic decays into charmed meson doublet ($0^+$,$1^+$) in HQET can be found
in the literatures by using various approaches, including HQET-based
considerations \cite{IWsr,Leib}, QCD sum rules \cite{c-sum,h-dai,h-dai01} and
various quark models \cite{wambach,veseli,oliver,DDG,dai2,EFG}. In this work, we
shall calculate the weak decay elements for $B_s\to D_{s0}^*,D_{s1}^*$ in HQET.
Considering that most of the phase space for these decays is near zero recoil, we
shall include the $\Lambda_{\mathrm QCD}/m_Q$ corrections in the application.

The paper is organized as follows. In Sec. \ref{sec2} we review the formulas for
the matrix elements of the weak currents including the structure of the
$\Lambda_{\mathrm QCD}/m_Q$ corrections in HQET. The QCD sum rule analysis for
the leading and subleading Isgur-Wise functions is presented in Sec. \ref{sec3}.
Section \ref{sec4} is devoted to numerical analysis and the applications to decay
widths. This section also includes a brief summary.

\section{ Decay matrix elements and the heavy quark expansion}
\label{sec2}

The matrix elements of vector and axial vector currents ($V^\mu=\bar
c\,\gamma^\mu\,b$ and $A^\mu=\bar c\,\gamma^\mu\gamma_5\,b$) between $B_s$ meson
and excited $D_{s0}^*$ or  $D_{s1}^*$ mesons can be parameterized as
\begin{eqnarray}\label{matrix1}
\langle D_{s0}^*(v')|\, V^\mu\,|B_s(v)\rangle
  &=& 0\,,\nonumber\\*
{\langle D_{s0}^*(v')|\,A^\mu\,|B_s(v)\rangle}
  &=& g_+ (v^\mu+v'^\mu) + g_- (v^\mu-v'^\mu)\,, \nonumber\\*
{\langle D_{s1}^*(v',\epsilon)| \, V^\mu\,|B_s(v)\rangle}
  &=&g_{V_1} \epsilon^{* \mu}
  + (g_{V_2} v^\mu + g_{V_3} v'^\mu)\,
  \epsilon^*_{\alpha\beta}\, v^\alpha v^\beta \,,\nonumber\\*
{\langle D_{s1}^*(v',\epsilon)|\, A^\mu\, |B_s(v)\rangle}
  &=& i g_A \varepsilon^{\mu\alpha\beta\gamma}
  \epsilon^*_\alpha v_\beta v'_\gamma \,.
\end{eqnarray}
Here form factors $g_i$ are functions of the dot-product, $y=v\cdot v'$, of the
initial and final meson four-velocities. The differential decay rates expressed
in terms of the form factors are given by
\begin{subequations}
\begin{eqnarray}
   \frac{{\mathrm d}\Gamma_{D_{s0}^*}}{{\mathrm d}y}
   &=& \frac{G_F^2\,|V_{cb}|^2 m_B^5}{48\pi^3}\,r_0^3 \,(y^2-1)^{3/2}\,
  \Big[(1+r_0)g_+-(1-r_0)g_- \Big]^2  \,, \label{drate1}\\
   \frac{{\mathrm d}\Gamma_{D^*_{s1}}}
    {{\mathrm d}y}
   &=& \frac{G_F^2\,|V_{cb}|^2 m_B^5}{48\pi^3}\,r_1^3\,\sqrt{y^2-1}
   \bigg\{ \Big[ (y-r_1)\;g_{V_1}+(y^2-1)(g_{V_3}+r_1g_{V_2}) \Big]^2 \nonumber\\*
&& + 2(1-2r_1y+r_1^2)\Big[g_{V_1}^2+(y^2-1)g_A^2 \Big] \bigg\}
\,.\label{drate2}
\end{eqnarray}\end{subequations}
where $r_0=m_{D_{s0}^*}/m_{B_s}$ and $r_1=m_{D_{s1}^*}/m_{B_s}$.

The form factors $g_i$ can be expressed by a set of Isgur-Wise functions at each
order in $\Lambda_{\rm QCD}/m_{c,b}$. This is achieved by evaluating the matrix
elements of the effective current operators arising from the HQET expansion of
the weak currents. This problem has been discussed previously in Ref.
\cite{Leib}, we outline here the analysis for the sake of completeness.

One introduces the matrix representations
\begin{eqnarray}\label{c-rep}
&&H_v = \frac{1+\rlap/ v}2\, \Big[ P_v^{*\mu} \gamma_\mu
  - P_v\, \gamma_5 \Big] \,,\nonumber\\
&&K_v = \frac{1+\rlap/ v}2\, \Big[ P_v^{'*\mu}\, \gamma_5\gamma_\mu
  + P'_{v}\, \Big]
\end{eqnarray}
composed from the fields $P_v$, $P_v^{*\mu}$ and $P'_v$, $P_v^{'\mu}$ that
destroy mesons in the doublets $j_\ell^P=\frac{1}{2}^-$ and $\frac{1}{2}^+$ with
four-velocity $v$ in HQET, respectively. At leading order of the heavy quark
expansion the hadronic matrix elements of weak current between the states
annihilated by the fields in $H_v$ and $K_{v'}$ are written as
\begin{eqnarray}
\label{leading} \bar h^{(c)}_{v'}\, \Gamma\, h^{(b)}_v = \zeta(w)\,
  {\rm Tr}\, \Big\{ \bar K_{v'}\, \Gamma\, H_v \Big\} \,.
\end{eqnarray}
where $h_v^{(Q)}$ is the heavy quark field in the effective theory and $\zeta$ is
a universal Isgur-Wise function of $y$.

At order $\Lambda_{\rm QCD}/m_Q$ there are contributions to the decay matrix
elements originating from corrections to the HQET Lagrangian
 \begin{eqnarray}\label{dlagr}
\delta{\cal L}=\frac{1}{2 m_Q} \Big[ O_{{\rm kin},v}^{(Q)}+O_{{\rm
mag},v}^{(Q)} \Big]\;,\hspace{2.5cm}\nonumber\\
O_{{\rm kin},v}^{(Q)}=\bar h_v^{(Q)} (iD)^2 h_v^{(Q)}\;,\hspace{0.2cm}
O_{{\rm
mag},v}^{(Q)}=\bar h_v^{(Q)}\frac{g_s}2 \sigma_{\alpha\beta} G^{\alpha\beta}
h_v^{(Q)}
\end{eqnarray}
and originating from the matching of the $b\to c$ flavor changing current onto
those in the effective theory
\begin{equation}\label{HQETcurrent}
\bar c\, \Gamma\, b = \bar h_{v'}^{(c)}\,
  \bigg( \Gamma - \frac i{2m_c} \overleftarrow{\rlap/ D}\;\Gamma
  + \frac i{2m_b}\, \Gamma \overrightarrow{\rlap/ D}
  \bigg)\, h_v^{(b)} \,,
\end{equation}
where $D$ is the covariant derivative. The matrix elements of the later
operators can be parameterized as
\begin{eqnarray}\label{curr}
\bar h^{(c)}_{v'}\, i\overleftarrow D_{\!\lambda}\, \Gamma\, h^{(b)}_v &=&
  {\rm Tr}\, \Big\{ {\cal S}^{(c)}_{\lambda}\,
  \bar K_{v'}\, \Gamma\, H_v \Big\} \,, \nonumber\\*
\bar h^{(c)}_{v'}\, \Gamma\, i\overrightarrow D_{\!\lambda}\, h^{(b)}_v &=&
  {\rm Tr}\, \Big\{ {\cal S}^{(b)}_{\lambda}\,
  \bar K_{v'}\, \Gamma\, H_v \Big\} \,.
\end{eqnarray}
The most general decomposition for ${\cal S}^{(Q)}_{\lambda}$ is
 \begin{equation}\label{Sdef}
{\cal S}^{(Q)}_{\lambda} = \zeta_1^{(Q)} v_\lambda + \zeta_2^{(Q)}
  v'_\lambda + \zeta_3^{(Q)} \gamma_\lambda \,.
\end{equation}
The functions $\zeta_i$ depend on $y$ and have mass dimension one. The
translation invariance, $i\partial_\nu\,(\bar h_{v'}^{(c)}\,\Gamma\,h_v^{(b)})
=(\bar\Lambda v_\nu-\bar\Lambda'v'_\nu)\,\bar h_{v'}^{(c)}\,\Gamma\,h_v^{(b)}$,
and the motion equation for the heavy quark, $iv\cdot D h_v^{(Q)}=0$, result in
the relations between the form factors $\zeta^{(Q)}_i$ \cite{Leib}
\begin{eqnarray}\label{const3}
&&\zeta_1^{(c)} + \zeta_1^{(b)} = \bar\Lambda\, \zeta \,, \hspace{1.2cm}
  \zeta_2^{(c)} + \zeta_2^{(b)} = -\bar\Lambda'\, \zeta \,, \hspace{1.2cm}
  \zeta_3^{(c)} + \zeta_3^{(b)} = 0\,, \nonumber\\
&&\zeta_2^{(c)}=-{y\bar\Lambda'-\bar\Lambda\over y+1}\, \zeta
  - \zeta_1^{(c)} \,,\hspace{1.38cm}
\zeta_3^{(c)}={y\bar\Lambda'-\bar\Lambda\over y+1}\, \zeta
  - (y-1)\, \zeta_1^{(c)}\;,
\end{eqnarray}
where $\bar\Lambda (\bar\Lambda')=m_M(m_{M'})-m_Q$ is the difference between
heavy ground state $j_\ell^P=\frac{1}{2}^-$ (excited $j_\ell^P=\frac{1}{2}^+$)
meson and heavy quark masses in the $m_Q\to\infty$ limit. These relations show
that all corrections to the form factors coming from the matching of the weak
currents in QCD onto those in the effective theory are expressible in terms of
$\zeta$ and $\zeta_1^{(c)}$.

The matrix elements of $\Lambda_{\rm QCD}/m_Q$ corrections from the insertions of
the kinetic energy operator $O_{\rm kin}$ have the structure
\begin{eqnarray}\label{corr-lag}
i \int {\rm d}^4x\, T\,\Big\{ O_{{\rm kin},v'}^{c}(x)\,
  \Big[ \bar h_{v'}^{(c)}\, \Gamma\, h_{v}^{(b)} \Big](0)\, \Big\}
  &=& \chi^{c}_{\rm ke}\, {\rm Tr}\, \Big\{\bar K_{v'}\, \Gamma\, H_v \Big\} \,,\nonumber \\
i \int {\rm d}^4x\, T\,\Big\{ O_{{\rm kin},v}^{(b)}(x)\,
  \Big[ \bar h_{v'}^{(c)}\, \Gamma\, h_{v}^{(b)} \Big](0)\, \Big\}
  &=& \chi^{b}_{\rm ke}\, {\rm Tr}\, \Big\{\bar K_{v'}\, \Gamma\, H_v \Big\}\,.
\end{eqnarray}
The functions $\chi^{c,b}_{\rm ke}(y)$ have mass dimension and effectively
correct the leading order Isgur-Wise function $\zeta(y)$ since the kinetic energy
operator does not violate heavy quark spin symmetry.

There are $\Lambda_{\mathrm QCD}/m_Q$ corrections associated with the insertion
of chromomagnetic operator $O_{\rm mag}$. The QCD sum rule approach for the
semileptonic $B$ decays to ground state and excited $D$ mesons shows that the
functions parameterizing the time-ordered products of the chromomagnetic term in
the HQET Lagrangian with the leading order currents are negligibly small
\cite{neubert-m,h-dai01}. This is in agreement with the results obtained from the
HQET-motivated considerations \cite{Leib} and relativistic quark model
\cite{EFG}. Therefore, we shall neglect the chromomagnetic correction hereafter.

Summing up all the contributions the resulting structure of the decay form
factors is
\begin{eqnarray}
g_+ &=& \varepsilon_c\, \bigg[ 2(y-1)\zeta_1
  - 3\zeta\, {y\bar\Lambda'-\bar\Lambda\over y+1} \bigg]
  - \varepsilon_b\, \bigg[ {\bar\Lambda'(2y+1)-\bar\Lambda(y+2)\over y+1}\,
  \zeta - 2(y-1)\,\zeta_1 \bigg] \,, \nonumber\\*
g_- &=& \zeta + \varepsilon_c\,  \chi^c_{\rm ke}
  + \varepsilon_b\, \chi^b_{\rm ke} \,,\nonumber\\
g_A &=& \zeta
  + \varepsilon_c\, \bigg[ {y\bar\Lambda'-\bar\Lambda \over y+1} \zeta
  +\chi^c_{\rm ke}\bigg] -
  \varepsilon_b\, \bigg[ {\bar\Lambda'(2y+1)-\bar\Lambda(y+2)\over y+1}\,
  \zeta - 2(y-1)\,\zeta_1 - \chi^b_{\rm ke} \bigg] \,,\nonumber\\*
g_{V_1} &=&  (y-1)\,\zeta + \varepsilon_c\,
  \Big[(y\bar\Lambda'-\bar\Lambda)\zeta + (y-1)\chi^c_{\rm ke} \Big]
  \nonumber\\*
&& - \varepsilon_b\, \Big\{ [\bar\Lambda'(2y+1)-\bar\Lambda(y+2)]\, \zeta
  - 2(y^2-1)\,\zeta_1 - (y-1)\chi^b_{\rm ke} \Big\} \,, \nonumber\\*
g_{V_2}&=& 2\varepsilon_c\, \zeta_1 \,,  \nonumber\\*
g_{V_3}&=&-\zeta-\varepsilon_c\, \bigg[ {y\bar\Lambda'-\bar\Lambda \over
y+1}\zeta+ 2\zeta_1 + \chi^c_{\rm ke}\bigg]+  \nonumber\\*&&\varepsilon_b\,
\bigg[ {\bar\Lambda'(2y+1)-\bar\Lambda(y+2)\over y+1}\,
  \zeta - 2(y-1)\,\zeta_1 - \chi^b_{\rm ke} \bigg] \,.\label{gexp1}
\end{eqnarray}
where $\varepsilon_Q=1/(2m_Q)$ and the superscript on $\zeta_1^{(c)}$ is dropped.
In the following sections we shall employ the QCD sum rule approach to calculate
the leading and subleading Isgur-Wise functions.

\section{Form factors from QCD Sum Rules}
\label{sec3}
\subsection{Leading Isgur-Wise function $\zeta$}
\label{subsec1}

The proper interpolating current $J_{j,P,j_{\ell}}^{\alpha_1\cdots\alpha_j}$ for
a heavy mesonic state with the quantum number $j$, $P$, $j_{\ell}$ in HQET was
given in \cite{huang}. These currents were proved to satisfy the following
conditions
\begin{eqnarray}
\label{decay} \langle
0|J_{j,P,j_{\ell}}^{\alpha_1\cdots\alpha_j}(0)|j',P',j_{\ell}^{'}\rangle&=&
f_{Pj_l}\delta_{jj'}
\delta_{PP'}\delta_{j_{\ell}j_{\ell}^{'}}\eta^{\alpha_1\cdots\alpha_j}\;,\\
\label{corr} i\:\langle 0|T\left
(J_{j,P,j_{\ell}}^{\alpha_1\cdots\alpha_j}(x)J_{j',P',j_{\ell}'}^{\dag
\beta_1\cdots\beta_{j'}}(0)\right
)|0\rangle&=&\delta_{jj'}\delta_{PP'}\delta_{j_{\ell}j_{\ell}'}
(-1)^j\:{\cal S}\:g_t^{\alpha_1\beta_1}\cdots g_t^{\alpha_j\beta_j}\nonumber\\[2mm]&&\times\:
\int \,dt\delta(x-vt)\:\Pi_{P,j_{\ell}}(x)
\end{eqnarray}
in the $m_Q\to\infty$ limit. Where $\eta^{\alpha_1\cdots\alpha_j}$ is the
polarization tensor for the spin $j$ state,
$g^{\alpha\beta}_t=g^{\alpha\beta}-v^\alpha v^\beta$ is the transverse metric
tensor, ${\cal S}$ denotes symmetrizing the indices and subtracting the trace
terms separately in the sets $(\alpha_1\cdots\alpha_j)$ and
$(\beta_1\cdots\beta_{j})$, $f_{P,j_{\ell}}$ and $\Pi_{P,j_{\ell}}$ are
a constant and a function of $x$ respectively which depend only on $P$ and $%
j_{\ell}$.

The local interpolating current for creating $0^-$ pseudoscalar $B_s$ meson is
taken as
\begin{eqnarray}
J^{\dag}_{0,-,{1/2}}=\sqrt{\frac{1}{2}}\:\bar h_v\gamma_5s\;,\label{p-scalar}
\end{eqnarray}
and the local interpolating currents for creating $0^+$ and $1^+$ ($D_{s0}^*$,
$D_{s1}^*$) mesons are taken as
\begin{eqnarray}
J^{\dag}_{0,+,1/2}&=&\frac{1}{\sqrt{2}}\:\bar h_v(-i)\rlap/{D}_ts\;,\nonumber \\
J^{\dag\alpha}_{1,+,1/2}&=&\frac{1}{\sqrt{2}}\:\bar
h_v\gamma^5\gamma^{\alpha}_t(-i)\rlap/{ D}_ts\;,\label{curr-01}
\end{eqnarray}

In order to calculate this form factor using QCD sum rules, we follows the same
procedure as \cite{neubert} to study the analytic properties of the three-point
correlators
\begin{subequations}\label{3-lead}
\begin{eqnarray}
 i^2\int d^4xd^4ze^{i(k'\cdot x-k\cdot z)}\langle 0|T\big(
 J_{0,+,1/2}(x)\;{\cal J}^{\mu}_{A}(0)\;
 J^{\dagger}_{0,-,1/2}(z)\big)|0\rangle&=&\Xi(\omega,\omega',y){\cal L}
 ^{\mu}_{A}\;, \\
i^2\int d^4xd^4ze^{i(k'\cdot x-k\cdot z)}\langle 0|T\big(
 J^{\nu}_{1,+,1/2}(x)\;{\cal J}^{\mu}_{V,A}(0)\;
 J^{\dagger}_{0,-,1/2}(z)\big)|0\rangle&=&\Xi(\omega,\omega',y){\cal L}
 ^{\mu\nu}_{V,A}\;,
\end{eqnarray}
\end{subequations}
where ${\cal J}^{\mu}_{V}=\bar h(v')\gamma^\mu\,h(v)$ and ${\cal
J}^{\mu}_{A}=\bar h(v')\gamma^\mu\gamma_5\,h(v)$ are leading order vector and
axial vector currents, respectively. The variables $k$, $k'$ denote residual
``off-shell" momenta which are related to the momenta $p$ of the heavy quark in
the initial state and $p'$ in the final state by $k=p-m_Qv$, $k'=p'-m_{Q'}v'$,
respectively. The Lorentz structures, ${\cal L}_{V,A}$, have the forms
\begin{eqnarray}
{\cal L}^{\mu}_{A}=v^\mu-v^{'\mu}\;,\hspace{4mm}{\cal
L}^{\mu\nu}_{A}=-i\epsilon^{\mu\nu\alpha\beta}v_\alpha v'_\beta\;,\hspace{4mm}
{\cal L}^{\mu\nu}_{V}=(y-1)g_t^{\mu\nu}-v'^\mu v_t^\nu\;,\nonumber
\end{eqnarray}
where $v_t^\alpha=g_t^{\alpha\beta}v_\beta=v^\alpha-yv'^\alpha$.

The coefficient $\Xi(\omega,\omega',y)$ in (\ref{3-lead}) is an analytic scalar
function in the ``off-shell energies" $\omega=2v\cdot k$ and $\omega'=2v'\cdot
k'$ with discontinuities for positive values of these variables. By saturating
the double dispersion integrals for the correlators in (\ref{3-lead}) with
physical intermediate states in HQET, one finds the hadronic representation of
the correlator as following
 \begin{eqnarray}
\label{pole-lead} \Xi_{hadro}(\omega,\omega',y)={f_{-,{1\over
2}}f_{+,{1/2}}\zeta(y) \over (2\bar\Lambda-\omega- i\epsilon
)(2\bar\Lambda'-\omega'- i\epsilon)}+\mbox{higher resonances} \;.
\end{eqnarray}
As the result of equation (\ref{decay}), only one state with $j^P=1^+$
contributes to (\ref{pole-lead}), the other resonance with the same quantum
number $j^P$ and different $j_\ell$ does not contribute.

On the other hand, the correlator can be calculated in QCD in the Euclidean
region, i.e., for large negative values of $\omega$ and $\omega'$, in terms of
perturbative and nonperturbative contributions. Furthermore, the nonperturbative
effects are able to be encoded in vacuum expectation values of local operators,
the condensates. Hence one has
\begin{eqnarray}\label{theo}
\Pi(\omega, \omega^\prime, y) = \int d \nu d \nu^\prime {\rho^{pert}(\nu,
\nu^\prime, y) \over (\nu - \omega - i \epsilon) (\nu^\prime - \omega^\prime - i
\epsilon) }  + \Pi_{\rm cond}+ {\rm subtractions} \;.
\end{eqnarray}

The QCD sum rule is obtained by equating the phenomenological and theoretical
expressions for $\Xi$. In doing this the quark-hadron duality needs to be assumed
to model the contributions of higher resonance part of Eq. (\ref{pole-lead}).
Generally speaking, the duality is to simulate the resonance contribution by the
perturbative part above some threshold energies. In the QCD sum rule analysis for
$B$ semileptonic decays into ground state $D$ mesons, it is argued by Blok and
Shifman in \cite{shifman} that the perturbative and the hadronic spectral
densities can not be locally dual to each other, the necessary way to restore
duality  is to integrate the spectral densities over the ``off-diagonal''
variable $\omega_-=(\nu-\nu')/2$, keeping the ``diagonal'' variable
$\omega_+=(\nu+\nu')/2$ fixed. It is in $\omega_+$ that the quark-hadron duality
is assumed for the integrated spectral densities. We shall use the same
prescription in our application.

In order to suppress the contributions of higher resonance states a double Borel
transformation in $\omega$ and $\omega'$ is performed to both sides of the sum
rule, which introduces two Borel parameters $T_1$ and $T_2$. For simplicity we
shall take the two Borel parameters equal: $T_1 = T_2 =2T$.

The calculations of $\rho_{\rm pert}$ and $\Pi_{\rm cond}$ in HQET are
straightforward. In doing this, for simplicity, the residual momentum $k$ is
chosen to be parallel to $v$ such that $k_\mu=(k\cdot v)v_\mu$ (and similar for
$k'$) in the theoretical calculation. Since we deal with $\bar Qs$ states, the
light quark mass shall be included in calculating the spectral function and
condensates. For the condensates we confine us to the operators with dimension
$D\leq 5$ in OPE. By performing the Taylor expansion around $x_\mu=0$, we obtain
the expansion for the quark-condensate \cite{ESS}:
\begin{eqnarray}\label{qconds}
\langle :\bar\psi_\sigma(x)\overleftarrow{D}_\alpha\psi_\rho(y):\rangle&=&
\frac{\langle\bar qq\rangle}{ 16}[im_q(\gamma_\alpha)_{\rho\sigma}
-\delta_{\rho\sigma}m^2_q(x_\alpha-y_\alpha)] +\frac{\langle\bar q g_s\sigma\cdot
G q\rangle}{32}\big\{
(g_{\alpha\mu}-\frac{i}{3}\sigma_{\alpha\mu})_{\rho\sigma}\times \nonumber\\&&
(x^\mu-y^\mu)+ \frac{i}{12} m_q[(\gamma_\alpha)_{\rho\sigma}(x-y)^2+2(\rlap/
x-\rlap/
y)_{\rho\sigma}(x_\alpha-y_\alpha)]\nonumber\\&&-m_q^2\big[\frac{\delta_{\rho\sigma}}{6}
(x-y)^2(x_\alpha-y_\alpha)-\frac{i}{36} \sigma^{\mu\nu}_{\rho\sigma}((x-y)^2y_\mu
g_{\nu\alpha}\nonumber\\&& +2(x_\alpha-y_\alpha)y_\mu x_\nu)\big]\big\}\;.
\end{eqnarray}

After making double Borel transformations in the variables $\omega$ and $\omega'$
and changing the integral variables $\omega_+=(\nu+\nu')/2$,
$\displaystyle{\omega_-=(\frac{y+1}{y-1})^{1/2}(\nu-\nu')/2}$, one obtains the
sum rule for $\zeta$ as follows
\begin{eqnarray}
\label{zeta-sr} \zeta(y)\;f_{-,\frac{1}{2}}\,f_{+,\frac{1}{2}}\;e^{-(\bar\Lambda
+\bar\Lambda')/T}=
\frac{1}{8\pi^2}\frac{1}{(y+1)^2}\;\int_{2m_s}^{\omega_c}d{\omega_+}\,\,e^{-\omega_+/T}
\big[\omega_+^3+3(1+y)(m_s\omega_+^2+m_s^2\omega_+)\big]\nonumber\\*
-\frac{\langle\bar ss\rangle}{8}\big(3 m_s-m_s^2\frac{y+1}{T}\big)
-\frac{1}{12}m^2_0{\langle\bar ss\rangle}{1+y\over T}\big(1-{5m_s\over 8T}
+\frac{m_s^2}{48}\frac{15y+7}{T^2}\big)\hspace{0.25cm} \nonumber\\*
-\frac{1}{192}\langle\frac{\alpha_s}{\pi} GG\rangle\frac{y-1}{y+1}+
\frac{m_s}{32T}\langle\frac{\alpha_s}{\pi} GG\rangle\big(2\gamma_E-
\ln\frac{T^2}{\mu^2}+ \ln\frac{y+1}{2}\big)\;,\hspace{1.3cm}
\end{eqnarray}
where $m_0^2\,\langle\bar ss\rangle=\langle\bar
sg\sigma_{\mu\nu}G^{\mu\nu}s\rangle$ with $m_0^2=0.8 \mbox{GeV}^2$.

\subsection{Subleading Isgur-Wise function $\zeta_1$}
\label{subsec2}

In order to derive the QCD sum rule for the subleading form factor
$\zeta_1(y)$ defined in (\ref{Sdef}), we consider the following three-point
correlation functions
\begin{subequations}\label{3-point2}
\begin{eqnarray}
\Xi_{0A}^{\mu}(\omega,\omega',y)&=&i^2\int\, d^4xd^4z\,e^{i(k'\cdot x-k\cdot
z)}\;\langle 0|T\left(
 J_{0,+,1/2}(x)\;\bar h^{(c)}_{v'}\, i\overleftarrow{\rlap/ D}\gamma^\mu\gamma_5\,
 h^{(b)}_v(0)\;J^{\dagger}_{0,-,1/2}(z)\right)|0\rangle\nonumber\\&=&\Xi_{0}(\omega,\omega',y){\cal L}^{\mu}_{0A}\;,
\label{3-1a} \\
\Xi_{1A}^{\mu\nu}(\omega,\omega',y)&=&i^2\int\, d^4xd^4z\,e^{i(k'\cdot x-k\cdot
z)}\;\langle 0|T\left(
 J^{\nu}_{1,+,1/2}(x)\;\bar h^{(c)}_{v'}\, i\overleftarrow{\rlap/ D}\gamma^\mu\gamma_5\,
 h^{(b)}_v(0)\;J^{\dagger}_{0,-,1/2}(z)\right)|0\rangle\nonumber\\&=&\Xi_{1}(\omega,\omega',y){\cal L}^{\mu\nu}_{A}\;,
\label{3-2v}\\
\Xi_{1V}^{\mu\nu}(\omega,\omega',y)&=&i^2\int\, d^4xd^4z\,e^{i(k'\cdot x-k\cdot
z)}\;\langle 0|T\left(
 J^{\nu}_{1,+,1/2}(x)\;\bar h^{(c)}_{v'}\, i\overleftarrow{\rlap/ D}\gamma^\mu\,
h^{(b)}_v(0)\;J^{\dagger}_{0,-,1/2}(z)\right)|0\rangle\nonumber\\&=&\Xi_{2}(\omega,\omega',y){\cal
L}^{\mu\nu}_{\zeta V}+\Xi_{3}(\omega,\omega',y){\cal L}^{\mu\nu}_{\zeta_1 V}\;,
\label{3-2a}
\end{eqnarray}
\end{subequations}
where the Lorentz structures have the forms
\begin{eqnarray}
{\cal L}_{0A}^\mu=v^\mu+v^{'\mu}\;, \hspace{1.2cm}{\cal L}^{\mu\nu}_{\zeta
V}=g^{\mu\nu}_t(y+1)-v^{'\mu}v^\nu_t\;,\hspace{1.2cm}{\cal L}^{\mu\nu}_{\zeta_1
V}=v^\mu v^\nu_t-v^{'\mu}v^\nu_t\;.\nonumber
\end{eqnarray}
The coefficient functions $\Xi_i(\omega,\omega',y)$ can be expressed in terms
of perturbative and nonperturbative contributions in QCD theoretical
calculation. These functions are used to construct the sum rules needed.

By saturating the double dispersion integral for the three-point functions in
(\ref{3-point2}) with  hadron states and using Eqs. (\ref{curr})-(\ref{const3})
and (\ref{decay}), one can isolate the contributions from the double pole at
$\omega=2\bar\Lambda$, $\omega'=2\bar\Lambda'$:
\begin{subequations}\label{Xi-pole}
\begin{eqnarray}
\Xi_{0A}^{\mu}(\omega,\omega',y)&=&{f_{-,{1\over 2}}f_{+,1/2}\over
(2\bar\Lambda-\omega- i\epsilon )(2\bar\Lambda'-\omega'-
i\epsilon)}\big[\,2(y-1)\zeta_1(y)-\nonumber\\&&-3{y\bar\Lambda' -
\bar\Lambda\over y+1}\zeta(y)\big]\,{\cal L}_{0A}^\mu+\cdots\;\;,\label{Xi1-pole1}\\
\Xi_{1A}^{\mu\nu}(\omega,\omega',y)&=&{f_{-,{1\over 2}}f_{+,{1/2}}\over
(2\bar\Lambda-\omega- i\epsilon )(2\bar\Lambda'-\omega'-
i\epsilon)}{y\bar\Lambda'-\bar\Lambda\over y+1}\zeta(y)
{\cal L}^{\mu\nu}_{A}+\cdots\;\;,\label{Xi1-pole2}\\
\Xi_{1V}^{\mu\nu}(\omega,\omega',y)&=&{f_{-,{1\over 2}}f_{+,{1/2}}
 \over (2\bar\Lambda-\omega-
i\epsilon)(2\bar\Lambda'-\omega'- i\epsilon)}[-{y\bar\Lambda' - \bar\Lambda\over
y+1} \zeta(y){\cal L}^{\mu\nu}_{\zeta V}\nonumber\\&&-2\zeta_1(y){\cal
L}^{\mu\nu}_{\zeta_1 V}]+\cdots\;\;.\label{Xi1-pole3}
\end{eqnarray}
\end{subequations}
From (\ref{3-point2}) and (\ref{Xi-pole}) one can see that in the case of
$\Xi_3$, the residue of the pole is proportional to the universal function
$\zeta_1(y)$, the pole contribution to $\Xi_1$ and $\Xi_2$ is related to
$\zeta(y)$. While for $\Xi_0$ the residue of the pole is proportional to both
$\zeta(y)$ and  $\zeta_1(y)$. QCD sum rule is obtained by equating the
phenomenological and theoretical expressions for $\Xi$. Therefore, the sum rule
for the subleading form factor $\zeta_1(y)$ can be constructed either from
$\Xi_3$ or from $\Xi_0$. One can also yield the sum rule for the leading form
factor in the form $(y\bar\Lambda'-\bar\Lambda)\zeta(y)$ from $\Xi_0$, $\Xi_1$
and $\Xi_2$, respectively.

We shall focus on the coefficient function $\Xi_3(\omega,\omega',y)$ to construct
the sum rule for the subleading form factors $\zeta_1(y)$. One obtains the sum
rule for $\zeta_1(y)$ as follows
\begin{eqnarray}\label{zeta1-sr}
\zeta_1(y)\,f_{-,\frac{1}{2}}\,f_{+,\frac{1}{2}}\;e^{-(\bar\Lambda
+\bar\Lambda')/T}&=&
-\frac{1}{16\pi^2}\frac{1}{(y+1)^3}\;\int_{2m_s}^{\omega_c}d{\omega_+}e^{-\omega_+/T}\,\big[
\omega_+^4+4m_s(y+1)(2y+1)^2\omega_+^3\big]\,\nonumber\\ &&
+\frac{5}{96}\;m_sm_0^2 {\langle\bar ss\rangle\over T}-\frac{1}{96} \langle
\frac{\alpha_s}{\pi}GG\rangle\frac{y}{(y+1)^2}\;T \;.
\end{eqnarray}
Moreover, the sum rule for the combination $(y\bar\Lambda'-\bar\Lambda)\zeta(y)$
can be obtained independently from the coefficient function $\Xi_1$ and $\Xi_2$
in (\ref{3-point2}) together with (\ref{Xi-pole}), respectively. We have double
checked that the resulted sum rule for $\zeta(y)$ has the same form as
(\ref{zeta-sr}). The above consistency checks confirm that our method is
consistent with the general analysis of Ref. \cite{Leib} described in Sec.
\ref{sec2}.

\subsection{QCD sum rules for  $\eta^{c,b}_{ke}$}
\label{subsec3}
For the determination of the form factor $\eta^{c}_{\rm ke}$, which relates to
the insertion of $\Lambda_{\rm QCD}/m_c$ kinetic operator of the HQET Lagrangian,
one studies the analytic properties of the three-point correlators
\begin{subequations}\label{3-point3}
\begin{eqnarray}
 i^2\int\, d^4xd^4x'd^4z\,e^{i(k'\cdot x'-k\cdot x)}\;\langle 0|T\left(
 J_{0,+,1/2}(x')\;O_{{\rm kin},v'}^{(c)}(z)\,{\cal J}^{\mu}_{A}(0)\;
 J^{\dagger}_{0,-,1/2}(x)\right)|0\rangle \nonumber\\=
 \Xi(\omega,\omega',y)\;{\cal L}^{\mu}_{A}\;, \hspace{3cm}\\
i^2\int\, d^4xd^4x'd^4z\,e^{i(k'\cdot x'-k\cdot x)}\;\langle
0|T\left(
 J^{\nu}_{1,+,1/2}(x')\;O_{{\rm kin},v'}^{(c)}(z)\,{\cal J}^{\mu}_{V,A}(0)\;
 J^{\dagger}_{0,-,1/2}(x)\right)|0\rangle\nonumber\\=
 \Xi(\omega,\omega',y)\;{\cal L}^{\mu\nu}_{V,A}\;.\hspace{3cm}
\end{eqnarray}
\end{subequations}

By saturating (\ref{3-point3}) with physical intermediate states in HQET, one can
isolate the contribution of interest as the one having poles at
$\omega=2\bar\Lambda$, $\omega'=2\bar\Lambda'$. Notice that the insertions of the
kinetic operator not only renormalize the leading Isgur-Wise function, but also
the meson coupling constants and the physical masses of the heavy mesons which
define the position of the poles. The correct hadronic representation of the
correlator is
\begin{eqnarray}
\label{pole1} \Xi_{\rm hadro}(\omega,\omega',y)&=&{f_{-,{1/2}}f_{+,{1/2}} \over
(2\bar\Lambda-\omega- i\epsilon )(2\bar\Lambda'-\omega'-
i\epsilon)}\;\bigg(\eta^c_{\rm ke}(y)+\nonumber\\&& (G_{+,1/2}^K+\frac{K_{+,1/2}}
{2\bar\Lambda'-\omega'- i\epsilon})\;\zeta(y)\bigg)+{\rm higher ~resonance} \;,
\end{eqnarray}
where $K_{P,j_\ell}$ and $G_{P,j_\ell}^K$ are defined by \cite{huang,dai-zhu}
\begin{eqnarray}
\langle j,P,j_\ell|O_{{\rm kin},v}^{(Q)}|j,P,j_\ell\rangle&=&K_{P,j_\ell}\;,\nonumber\\
\langle 0|i\int d^4x\;O_{{\rm
kin},v}^{(Q)}(x)J_{j,P,j_\ell}^{\alpha_1\cdots\alpha_j}(0) |j,P,j_\ell\rangle
&=&f_{P,j_\ell}\;G_{P,j_\ell}^K \eta^{\alpha_1\cdots\alpha_j}\;.\label{G-K}
\end{eqnarray}

Within the same procedure one finds the sum rule for $\eta^c_{\rm ke}$ as
\begin{eqnarray}\label{etak-sr}
\big[\eta^c_{\rm ke}(y)+ (G_{+,1/2}^K+\frac{K_{+,1/2}}{2T})\zeta(y)\big]
\,f_{-,1/2}f_{+,1/2}\; e^{-(\bar\Lambda+\bar\Lambda')/T} =
\hspace{3.5cm}\nonumber\\
-\frac{3}{16\pi^2}\frac{3y+2}{(y+1)^3}\;\int_{2m_s}^{\omega_c}d{\omega_+}\,e^{-\omega_+/T}
\big[\omega_+^4+\frac{4}{3}m_s(1+y)\omega_+^3\big]~~~~\nonumber\\+\frac{5}{32}\;m_sm_0^2
{\langle\bar ss\rangle\over T}+\frac{1}{96}\langle\frac{\alpha_s}{\pi}GG\rangle
\frac{3y+1}{(y+1)^2}T\;,\hspace{3cm}
\end{eqnarray}

From the consideration of symmetry, the sum rule for $\eta_{\rm ke}^{b}$ that
originates from the insertion of $\Lambda_{\rm QCD}/m_b$ kinetic operator of the
HQET Lagrangian is of the same form as in (\ref{etak-sr}), but with the HQET
parameters $G_{+,1/2}^{K}$ and $K_{+,1/2}$ replaced by $G_{-,1/2}^{K}$ and
$K_{-,1/2}$, respectively. The definitions of $G_{-,1/2}^{K}$ and $K_{-,1/2}$ can
be found in Eq. (\ref{G-K}).

We end this subsection by noting that the QCD $O(\alpha_s)$ corrections have not
been included in the sum rule calculations. However, the Isgur-Wise functions
obtained from the QCD sum rule actually are a ratio of the three-point correlator
to the two-point correlator results.  While both of these correlators subject to
large perturbative QCD corrections, the remaining corrections to the form factors
themselves may in fact be small because of cancellation. This is what happens in
the case of analysis for $B$ semileptonic decay to ground state and excited
charmed mesons \cite{neubert-r,CDP}.

\section{Numerical results and predictions for semileptonic decay widths}
\label{sec4}

We now turn to the numerical evaluation of these sum rules and the
phenomenological implications. In order to obtain information for $\zeta(y)$,
$\zeta_1(y)$ and $\eta^{c,b}_{\rm ke}(y)$ from the sum rules,  we need the
related decay constants, $f_{-,1/2}$ and $f_{+,1/2}$, defined in (\ref{decay}) as
input. The QCD sum rule calculations for the correlators of two heavy-light
currents give \cite{huangt,DHLZ}:
\begin{eqnarray}
f_{-,1/2}^2\,e^{-2\bar\Lambda/T} &=&\frac{3}{16\pi^2}
\int_{2m_s}^{\omega_{0}}d\omega(\omega^2+2m_s\omega-2m_s^2)e^{-\omega/{T}}-
\frac12\langle\bar
ss\rangle\left(1-\frac{m_s}{2T}+\frac{m_s^2}{2T^2}\right)\nonumber\\&&
+\frac{m_0^2}{8 T^2}\langle\bar
ss\rangle\left(1-\frac{m_s}{3T}+\frac{m_s^2}{3T^2}\right)
-\frac{m_s}{16T^2}\langle\frac{\alpha_s}{\pi}GG\rangle\left(2\gamma_E-1-
\ln\frac{T^2}{\mu^2}\right),\label{2-point1}\\
{f'}_{+,1/2}^2e^{-2\bar\Lambda^\prime/{T}}&=&
\frac{3}{64\pi^2}\int_{2m_s}^{\omega_1} [\omega^4+2m_s \omega^3-6m_s^2
\omega^2-12 m_s^3 \omega] e^{-\omega/{T}}d\omega\nonumber
\\&&-\frac{1}{16}\,m_0^2\,\langle\bar ss\rangle(1-\frac{m_s}{T}+\frac43\frac{m_s^2}{T^2})
+{3\over 8}m_s^2 \,\langle\bar ss\rangle-{m_s\over 16\pi}\,\langle \alpha_s G^2
\rangle\;.\label{2-point2}
\end{eqnarray}
These two-point sum rules can be used to eliminate the explicit dependence of
three-point sum rules on $f_{-,1/2}$ and $f_{+,1/2}$, as well as on $\bar\Lambda$
and $\bar\Lambda'$. This procedure may help to reduce the uncertainties in the
calculation.

For the QCD parameters entering the theoretical expressions, we take the standard
values
 \begin{eqnarray}\label{cond}
\langle\bar q q\rangle &=& -(0.24\pm 0.01)^3~\mbox{GeV}^3\,, \nonumber\\
\langle{\alpha_s\over\pi} GG\rangle &=& (0.012\pm0.004)~\mbox{GeV}^4 \,.
\end{eqnarray}
The strange/nonstrange condensate ratio is adopted as $\langle\bar s
s\rangle=(0.8\pm 0.1) \langle\bar q q\rangle$. The strange quark mass is taken as
$m_s(1 {\rm GeV})=0.15$ GeV, and the cut-off parameter is chosen as $\mu=1$ GeV.

Let us evaluate numerically the sum rules  for $\zeta(y)$ and $\zeta_1(y)$. The
continuum thresholds $\omega_{0}$ and $\omega_{1}$ in (\ref{2-point1}) and
(\ref{2-point2}) are determined by requiring stability of these sum rules. One
finds that $1.7~{\rm GeV}<\omega_{0}<2.2~{\rm GeV}$ and $2.6~{\rm
GeV}<\omega_{1}<3.1~{\rm GeV}$ \cite{neubert,huangt,DHLZ}. Imposing usual
criterion on the ratio of contribution of the higher-order power corrections and
that of the continuum, we find that for the central values of the condensates
given in (\ref{cond}), if the threshold parameter $\omega_c$ lies in the range
$2.3<\omega_c<2.7$ GeV, there is an acceptable ``stability window'' $T=0.8-1.2$
GeV in which the calculation results do not change appreciably.

\begin{figure}[t]
\centerline{\epsfysize=5truecm \epsfbox{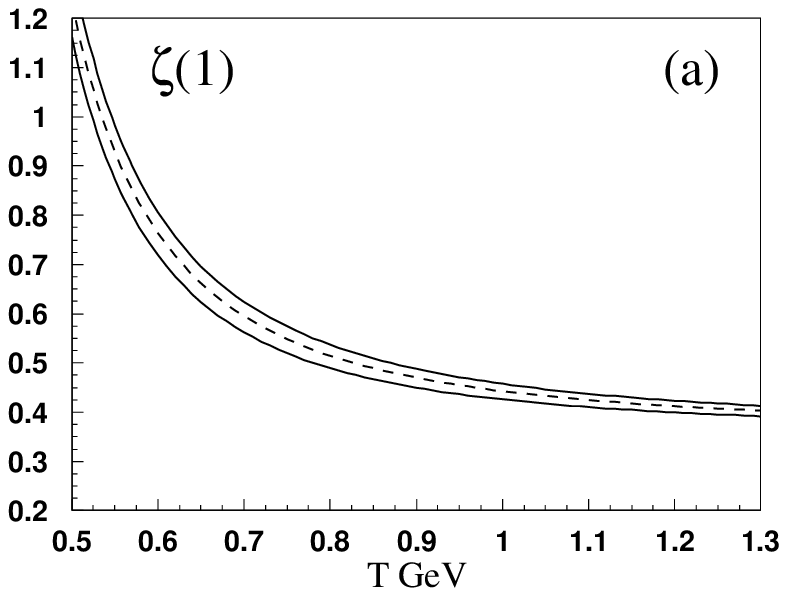} \epsfysize=5truecm
\epsfbox{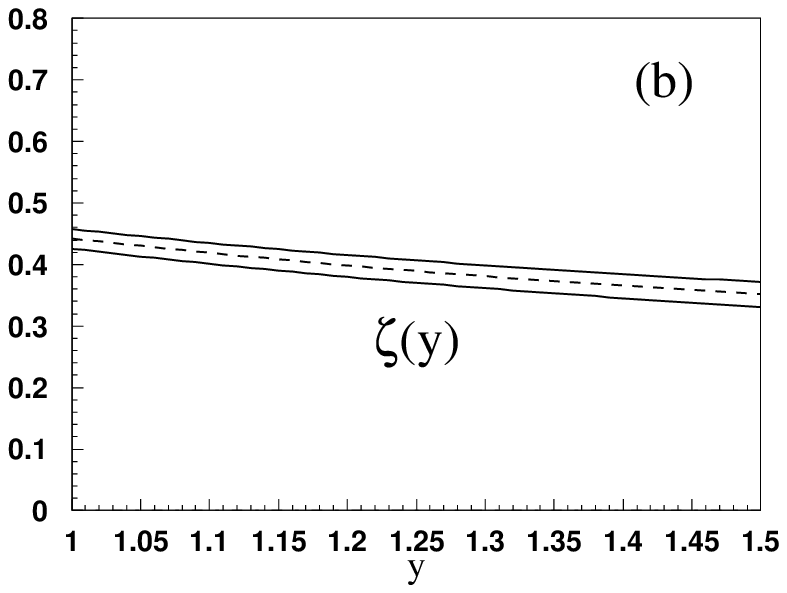}} \caption{Fig. 1(a) shows the dependence of $\zeta(1)$ on
the Borel parameter $T$ for the continuum threshold in the range
$2.3<\omega_c<2.7$ GeV; Fig. 1(b) shows Isgur-Wise function $\zeta(y)$ with
$T=1.0$ GeV.} \label{fig:1}
\end{figure}

The values of the form factors $\zeta(y)$ and $\zeta_1(y)$ at zero recoil as
functions of the Borel parameter are shown in Fig. 1(a) and 2(a), for three
different values of the continuum threshold $\omega_c$. One can see that the
variation is quit moderate in the range $0.8<T<1.2$ GeV. The numerical results
for $\zeta(y)$ and $\zeta_1(y)$ are shown in Fig. 1(b) and 2(b), where the curves
refer to three different values of $\omega_c$ and $T$ is fixed at $T=1.0$ GeV.

\begin{figure}[b]
\centerline{\epsfysize=5truecm \epsfbox{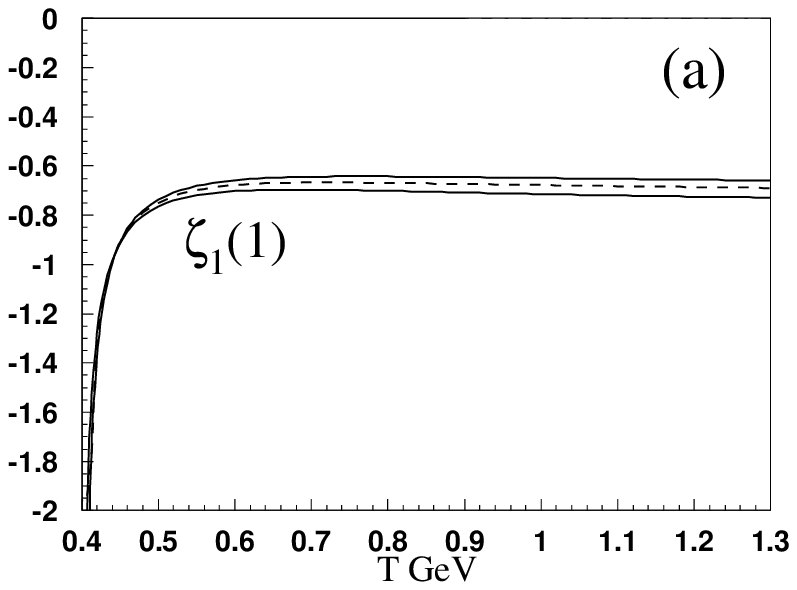}
  \epsfysize=5truecm \epsfbox{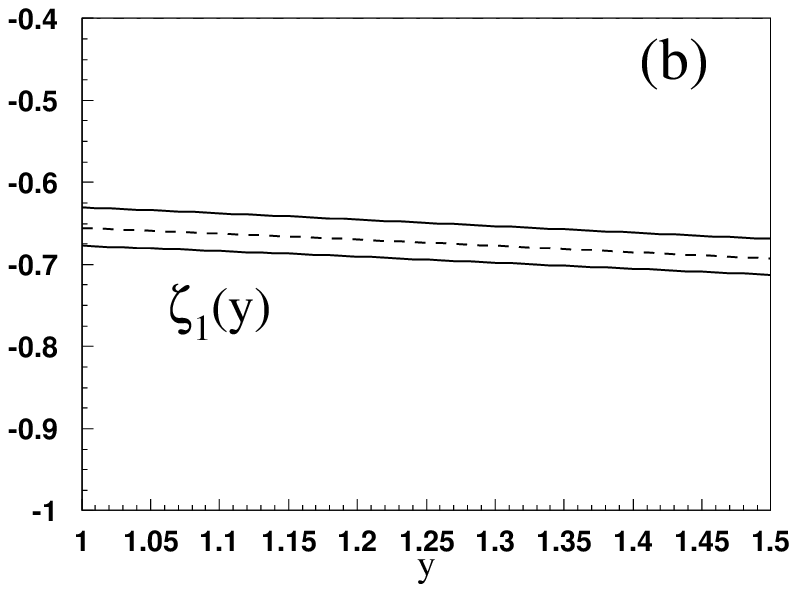}}
\caption{Numerical result for the sum rule (\ref{zeta1-sr}): (a) dependence of
$\zeta_1(1)$ on the Borel parameter $T$ for the continuum threshold in the range
$2.3<\omega_c<2.7$ GeV; (b) The form factor $\zeta_1(y)$ with $T=1.0$ GeV.}
\label{fig:2}
\end{figure}

In order to evaluate numerically the sum rules for $\eta^c_{\rm ke}(y)$ and
$\eta_{\rm ke}^b(y)$, we need to specify the following HQET parameters as input,
which are obtained by QCD sum rules \cite{DHLZ,dai-zhu,bball}:
\begin{eqnarray}
&&K_{+,3/2}=-(1.6\pm0.30)~{\rm GeV}^2\;,\hspace{1.0cm}
G^K_{+,3/2}=-(1.0\pm0.45)~{\rm GeV}\nonumber\\ &&K_{-,1/2}=-(1.2\pm0.20)~{\rm
GeV}^2\;, \hspace{1.cm}G^K_{-,1/2}=-(1.6\pm0.6)~{\rm GeV}\;. \label{GK-numerical}
\end{eqnarray}
In Fig. 3(a), the sum rule for $\eta^c_{\rm ke}(y)$ is plotted at zero recoil as
a function of Borel parameter for various choices of the continuum thresholds in
the range $2.5<\omega_c<2.9$. Fig. 3(b) shows the $y$ dependence of the form
factor $\eta^c_{\rm ke}(y)$ for the central value of HQET parameters and $T=1.2$
GeV. It should be noted that apart from the uncertainty from the sum rule working
window, there is uncertainty to a large extent due to the variation of $K'$s and
$G_K'$s in the numerical analysis. The numerical evaluation for the sum rule of
$\eta^b_{\rm ke}(y)$ can follow the same procedure.

\begin{figure}[htb]
\centerline{\epsfysize=5truecm \epsfbox{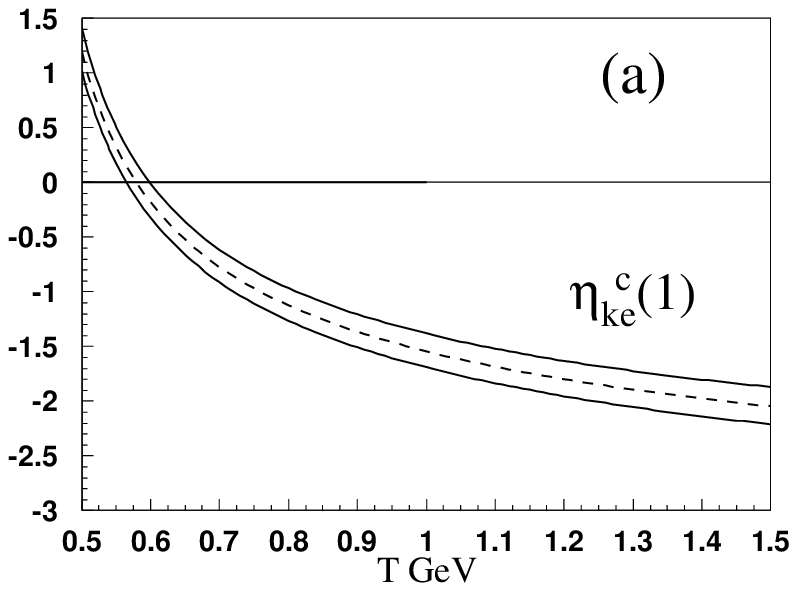} \epsfysize=5truecm
\epsfbox{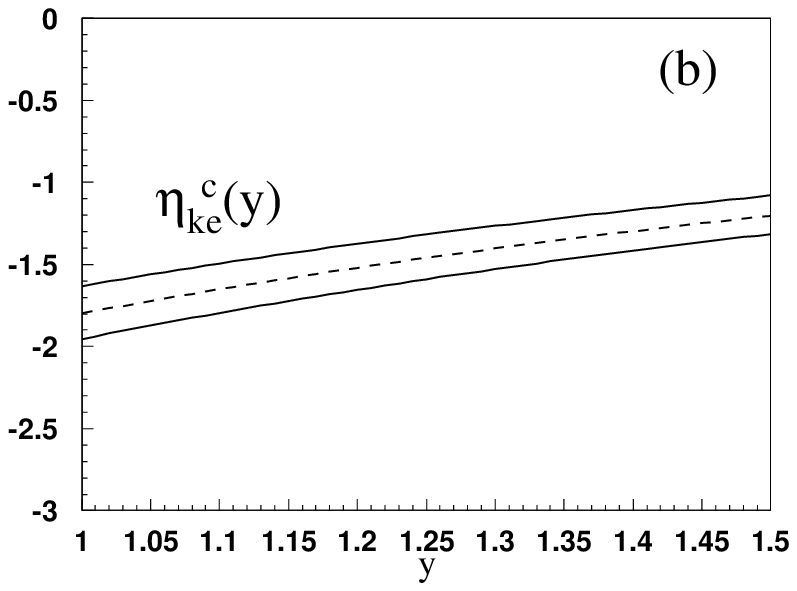}} \caption{Results of the numerical evaluation for the sum
rules (\ref{etak-sr}). The curves refer to choices of the threshold parameters:
$\omega_c=2.5$, $\omega_c=2.7$, $\omega_c=2.9$, from top to bottom.}
\label{fig:3}
\end{figure}

The numerical analysis shows that all Isgur-Wise functions $\zeta(y)$,
$\zeta_1(y)$ and $\eta^{c,b}_{\rm ke}(y)$ are slowly varying functions in the
allowed kinematic range for $B_s\to D_{s0}^*\ell\bar\nu$ and $B_s\to
D_{s1}^*\ell\bar\nu$ decays. They can be well fitted by the linear approximation
\begin{eqnarray}\label{value1}
\zeta(y)&=&\zeta(1)\big(1-0.5(y-1)\big)\;,\hspace{9mm} \zeta(1)=0.45\pm
0.05\;\;{\rm GeV} \nonumber\\
\zeta_1(y)&=&-\zeta_1(1)\big(1+0.4(y-1)\big)\;, \hspace{5mm} \zeta(1)=0.65\pm
0.06\;\;{\rm
GeV}\nonumber\\
\eta^c_{\rm ke}(y)&=&-\eta^c_{\rm ke}(1)\big(1-0.9(y-1)\big)\;,
\hspace{3mm}\eta^c_{\rm ke}(1)=1.7\pm 0.2 \hspace{0.1cm}{\rm
GeV} \;,\nonumber\\
\eta_{\rm ke}^b(y)&=&-\eta^b_{\rm ke}(1)\big(1-0.9(y-1)\big)\;,
\hspace{3mm}\eta_{\rm ke}^b(1)=1.6\pm 0.2 \hspace{0.1cm} {\rm GeV}\;.
\end{eqnarray}
The errors reflect the uncertainty due to $\omega$'s and $T$. The uncertainty due
to the variation of the QCD and HQET parameters is not included, which may reach
$5\%$ or more. The systematic error resulted from the use of quark-hadron duality
above $\omega_c$ is difficult to estimate. Conservatively speaking, there is a
$10\%$ systematic error.

Above parameterizations of the Isgur-Wise functions can be used to calculate the
total semileptonic rates and decay branching ratios by integrating Eqs.
(\ref{drate1}) and (\ref{drate2}). The values of $\bar\Lambda$ and $\bar\Lambda'$
can be obtained from two-point sum rules (\ref{2-point1}) and (\ref{2-point2}),
respectively. They are: $\bar\Lambda=0.62$ GeV \cite{huangt} and
$\bar\Lambda'=0.86$ GeV \cite{DHLZ}. The quark masses are taken to be $m_b=4.7$
GeV, $m_c=1.4$ GeV. We use the physical masses, $m_{B_s}=5.369$ \cite{PDG},
$m_{D_{s0}^*}=2.317$ and $m_{D_{s1}^\ast}=2.457$ \cite{Belle}, for $B_s$,
$D_{s0}^*$ and $D^*_{s1}$ mesons. The maximal values of $y$ in the present case
are $y_{0max}=(1+r_0^2)/2r_0=1.374$ and $y_{1max}=(1+r_1^2)/2 r_1=1.321$.

In Table \ref{tab:branch} we present our results for decay rates and branching
ratios, as well as those in the infinitely heavy quark limit. We have taken
$\tau_{B_s}=1.46$ ps \cite{PDG}. In the calculation, the central values for the
Isgur-Wise functions in (\ref{value1}) are taken, and the theoretical
uncertainties are not included. The ratios of the two semileptonic rates for
$B_s$ decays into $D_{s0}^0$ and $D_{s1}^*$ mesons both in taking account of the
$1/m_Q$ corrections and in the infinitely heavy quark mass limit are
\begin{equation}
R_{Br} \equiv {{\cal B}(B_s\to D_{s0}^*\ell\bar\nu) \over {\cal B}(B_s\to
D_{s1}^0\, \ell\bar\nu) } =\bigg\{\begin{array}{c@{\hspace{3ex}} c}2.05&
{\mbox{with~}} 1/m_Q\;,\\1.16 & m_Q\to\infty\;. \end{array}
\end{equation}
The branching ratio for $B_s\to D_{s0}^*\ell\bar\nu$ decay exceeds the one for
$B_s\to D_{s1}^*\ell\bar\nu$ in both cases.

\begin{table}[htb]
\begin{tabular}{c@{\hspace{3ex}}c@{\hspace{2ex}}c@{\hspace{3ex}}c
@{\hspace{2ex}}c@{\hspace{3ex}}c} \hline\hline
&\multicolumn{2}{c@{\hspace{3ex}}}{With $1/m_Q$}&
\multicolumn{2}{c@{\hspace{3ex}}}{$m_Q\to\infty$}\\
Decay& $\Gamma$ & Br& $\Gamma$ & Br &$R$ \\
\hline
$B\to D^*_{s1}\ell\bar\nu$&0.38&0.10 & 0.31 & 0.08&1.23 \\
$B\to D^*_{s0}\ell\bar\nu$&0.78&0.20 & 0.36 & 0.09&2.17 \\
\hline\hline
\end{tabular}
\caption{Decay rates $\Gamma$ (in units of $|V_{cb}/0.04|^2\times 10^{-15}$ GeV)
and branching ratios BR (in \%) for $B_s\to D_s^{(**)}\ell\bar\nu$ decays in
taking account of the $1/m_Q$ corrections and in the infinitely heavy quark mass
limit. $R$ is a ratio of branching ratios including ${\cal O}(1/m_Q)$ corrections
to branching ratios in the infinitely heavy quark mass limit.}\label{tab:branch}
\end{table}

The numerical predictions in Table \ref{tab:branch} indicate that a substantial
part of the inclusive semileptonic $B_s$ decays should go to excited $D_s$ meson
states. In future hadron $B$ factories the $D_s$ resonant states can be produced
directly in a considerable amount of branching ratio from the weak decay of the
$B_s$ meson. The study of the semileptonic $B_s$ decays to excited $D_s$ states
can provide some information about the structure and the properties for the newly
observed $D_{sJ}^*$ states.

From Table \ref{tab:branch} we see that the $B\to D_{s0}^*\ell\bar\nu$ decay rate
receives large $1/m_Q$ contributions and gets a sharp increase, while the $B\to
D^*_{s1}\ell\bar\nu$ decay rate is only moderately increased by subleading
$1/m_Q$ corrections. The reason for this is as following. From Eqs.
(\ref{matrix1}) and (\ref{gexp1}) we see that the decay matrix elements at zero
recoil are determined by form factors $g_+(1)$ and $g_{V_1}(1)$, which receive
non-vanishing contributions from first order heavy quark mass corrections.
Explicitly,
\begin{eqnarray}
g_+(1)&=&-\frac32(\varepsilon_c+\varepsilon_b)(\bar\Lambda'-\bar\Lambda)
\zeta(1)\;,\nonumber\\
g_{V_1}(1)&=&(\varepsilon_c-3\varepsilon_b)(\bar\Lambda'-\bar\Lambda)\zeta(1)\;.
\label{gat1}
\end{eqnarray}
At zero recoil the form factor $g_{V_1}$ is suppressed by a very small factor
$\varepsilon_c- 3\varepsilon_b\approx 0.04{\rm GeV}^{-1}$. As a result the
$B_s\to D^*_{s1}e\nu$ decay rate is only slightly increased by subleading $1/m_Q$
corrections. On the other hand, $B_s\to D^*_{s0}e\nu$ decay rate receives a large
enhancement from $1/m_Q$ corrections. Note that the sharp increase of $B_s\to
D^*_{s0}e\nu$ decay rate by $1/m_Q$ corrections does not imply the breakdown of
the heavy quark expansion. This is because the allowed kinematic ranges for $B\to
D_{S0}\ell\bar\nu$ is fairly small, the contribution to the decay rate of the
rather small $1/m_Q$ corrections is substantially increased. Hence it is rather a
result of kinematical and dynamical effects.

In summary, we have presented the investigation for semileptonic $B_s$ decays
into excited $D_s$ mesons. Within the framework of HQET we have applied the QCD
sum rules to calculate the universal Isgur-Wise functions up to the subleading
order of the heavy quark expansion. The differential decay widths and the
branching ratios for the decays $B_s\to D_{s0}^*\ell\bar\nu$ and $B_s\to
D_{s1}^*\ell\bar\nu$ are computed up to the order of $1/m_Q$ corrections. The
decay rates are substantially influenced by the inclusion of the first order
$1/m_Q$ corrections.

With the assumption that the newly discovered states $D^{*}_{sJ}(2317)$ and
$D^{*}_{sJ}(2460)$ can be identified with spin symmetry doublet
($D^*_{s0}$,$D^*_{s1}$), it is worthwhile to study the semileptonic $B_s$ decays
to these newly observed states. In hadron $B$ factories these $p$-wave excited
$D_s$ states can be produced directly from the semileptonic decay of the $B_s$
meson with a considerable amount of branching ratio. A measurement of the $B_s\to
D_{sJ}^*$ can provide some information on the nature of the new $D^{*}_{sJ}$
mesons.

\begin{acknowledgments}
I am very grateful to Y. B. Dai and C. S. Lam for their stimulating discussions
and useful suggestions. I would like to thank the McGill University for warm
hospitality. This work was supported in part by the National Natural Science
Foundation of China under Contract No. 10275091.
\end{acknowledgments}


\end{document}